\documentclass[twocolumn]{article}
\usepackage{graphicx}

\title{Measurement of Orthopositronium Decay Rate Using ${\rm SiO_2}$ Powder: 
Integration of Thermalization into Time Spectrum Fitting Procedure}
\author{O. Jinnouchi, S. Asai and T. Kobayashi \\ 
{\sl International Center for Elementary Particle Physics,}\\ 
{\sl University of Tokyo, Faculty of Science Building 1,}\\
{\sl 7-3-1 Hongo, Bunkyo-ku, Tokyo, 113-0033, Japan}}
\date{November, 2000}
\begin{document}
\twocolumn[
\maketitle
\centerline{\bfseries Abstract}\par
The intrinsic decay rate of orthopositronium (o-Ps) formed in  
${\rm SiO_2}$ powder was measured using a modified method which 
determined the time dependence of the pick-off annihilation rate 
using high-energy-resolution germanium detectors.
That is, the main systematic error due to thermalization was 
accounted for and integrated into the time spectrum fitting procedure. 
The averaged value was found to be 
$\lambda_{\mbox{\scriptsize o-Ps}}=7.0399^{\scriptscriptstyle +0.0020}_
{\scriptscriptstyle-0.0021}(stat.)\pm 0.0015(sys.)\mu s^{-1}$, 
which agrees well with the $O(\alpha^2)$ QED prediction; 
varying 2.8-4.2 experimental standard deviations from 
other measurements.
\vspace*{5mm}
]
%% macro 
\newcommand{\avee}{\overline{E(t)}}
\newcommand{\Tl}{Tl}

%-- INTRODUCTION --
%- theoretical background -
\section{Introduction}
\subsection{o-Ps lifetime problem}

The bound state of an electron and positron, positronium (Ps), 
is a pure quantum electrodynamical system that provides a highly 
sensitive field for testing accurate descriptions of bound states in 
quantum field theory. The triplet ($1^3S_1$) state of Ps, 
orthopositronium (o-Ps), will in most cases decay into three photons 
due to having odd-parity under the C-transformation. 
For the last two decades, vast number of theoretical works \cite{LOWEST-2,ADKINS-1,TH-LOG-0,TH-LLOG-0,ALPHA2-2,TH-LOG-2,ALPHA2-0,ALPHA2-1} have been done 
to yield the decay rate of o-Ps, up to some partial 
results on $O(\alpha^2)$ correction.   
Recent theoretical efforts have evaluated the value up to $O(\alpha^2)$ 
corrections (7.039~934(10)~$\mu s^{-1}$ \cite{ADKINS-4}), 
order $\alpha^3 \ln(1/\alpha)$ 
corrections (7.039~968(10)~$\mu s^{-1}$ \cite{LOG3-0,LOG3-1}), 
and so far neglected binding energy corrections \cite{ALPHA2-7} which is 
apart from \cite{ADKINS-4} by about $70~ppm$.
Therefore there still exists a $100~ppm$ level uncertainty on the 
theoretical side.
The decay lifetime is sufficient such that a large number of direct 
measurements have been carried out \cite{GIDLEY-78-GAS,GIDLEY-78-CAV,CANTER,LONDON,GIDLEY-82-GAS,MAINZ}. 
However recent precise measurements \cite{GAS87,GAS89,CAV90} have 
indicated that 
the measured decay rate is much larger, i.e., 1670, 1600, and 1170 ppm 
respectively, than the theoretical QED prediction \cite{ADKINS-4}.
These respective discrepancies of 9.1, 8.0, and 5.2 experimental 
standard deviations are statistically significant, being referred 
to as the `o-Ps lifetime problem.'

Although a variety of experiments have since been carried out to 
search for exotic decay modes of o-Ps \cite{EXOTIC-LL0,EXOTIC-LL1,EXOTIC-SL0,EXOTIC-SL1,EXOTIC-UB0,EXOTIC-IV0,EXOTIC-TW0}, none have
provided confirming evidence to elucidate this discrepancy.

\subsection{Thermalization of o-Ps}
We went on to perform further study \cite{ASAI94} in which two 
independent precision measurements were carried out using two types 
of ${\rm SiO_2}$ powder. Results correlated well, giving a 
combined value of
 \[\lambda_{\mbox{\scriptsize o-Ps}}=7.0398\pm0.0025(stat.)\pm0.0015(sys.)~\mu
s^{-1},\] which is consistent with the QED~$O(\alpha^2)$ prediction. 
This measurement employed a completely new method that took into account 
the thermalization process of o-Ps.

Since some fraction of o-Ps inevitably results in `pick-off' annihilations 
due to collisions with atomic electrons of the target material, the 
observed o-Ps decay rate $\lambda_{obs}$ is the sum of the intrinsic o-Ps 
decay rate $\lambda_{3\gamma}$ and the pick-off annihilation rate 
$\lambda_{pick}$, i.e.,
\begin{equation}
\lambda_{obs}(t) = \lambda_{3\gamma}+\lambda_{pick}(t),
\label{eq:rate_sum}
\end{equation} 
where $\lambda_{pick}$ is proportional to the collision rate of o-Ps 
with the target materials; hence being proportional to the densities 
of the target materials and the velocity of o-Ps.

Immediately after formation, o-Ps has a kinetic energy of about 1 $eV$, 
thermalizing via elastic collisions with surrounding molecules to an 
ultimate thermal energy of about 0.03 $eV$. 
Due to the deceleration or thermalization process, $\lambda_{pick}$ must 
be described as a function of time dependent on the properties of the 
respective materials. 
$\lambda_{obs}$ was previously measured \cite{GAS87,GAS89,CAV90} by 
varying the densities of the target materials, with the extrapolation 
to zero densities expected to yield $\lambda_{3\gamma}$ under the 
assumption of constant o-Ps velocity. 
Unfortunately, however, such extrapolation results in large systematic 
error due to an expected low thermalization rate, especially regarding 
the low density limit. Measurements in ${\rm SiO_2}$ powder indicated 
this systematic problem \cite{ASAI94,ASAI-D}; while in other experiments, 
o-Ps thermalization rates measured in gas were also substantially smaller than 
previously thought \cite{MICHIGAN-THERM}. 
Another problem is that the assumption of linear dependence of 
$\lambda_{pick}$ on the densities of the target materials is not 
always valid, i.e., o-Ps collisions with materials could very well 
be more complex due to the occurrence of multiple scattering at 
material surfaces.

Therefore, our previous measurements adopted a formulation containing 
all information representing the thermalization process \cite{ASAI94} ; 
hence completely removing systematic errors related to the 
thermalization process. 
From Eq. (\ref{eq:rate_sum}), the population of o-Ps, $N(t)$, can 
be expressed as
\begin{equation}
N(t)=N_0 \exp\left(-\lambda_{3\gamma}\int^t_0\left(1+\frac{\lambda_{pick}(t')}{\lambda_{3\gamma}}\right)dt'\right).
\label{eq:population}
\end{equation}
Note that the energy distribution of photons from the 
3-body $\gamma$-decay is continuous below the steep edge at 511 $keV$, 
whereas the final state of the pick-off 
annihilation is 2-body; a behavior producing the 511-$keV$ monochromatic 
peak. 
If the energy and timing information are simultaneously measured, 
$\lambda_{pick}(t)/\lambda_{3\gamma}$ can be determined from the 
energy spectrum of the emitted photon. Then, providing the ratio 
is determined as a function of time, the intrinsic o-Ps decay rate 
$\lambda_{3\gamma}$ can be directly obtained by fitting the observed 
time spectrum with Eq. (\ref{eq:population}); an approach which 
precludes any ambiguity due to the thermalization process.

%- experimental background -
\section{Experimental setup}
\subsection{Uncertainties in previous measurements} Although our 1995 
results were consistent with QED predictions, the error was large 
in comparison with that of other experiments \cite{GAS87,GAS89,CAV90}.
The time spectrum was obtained using CsI(Tl) scintillators which provided 
a relatively broad time response function ($\sigma\!\sim\!4~ns$) and long 
scintillation decay time. Accordingly, since some fraction of the prompt 
annihilation may have flown into the o-Ps decay region, uncertainties 
occurred in the time spectrum range immediately after the prompt events. 
This uncertainty led a restriction in the statistics such that we obtained 
a rather large error.

Here, although a new experimental setup/tech\-nique is used similar to the 
previous one, the above-mentioned problems are mitigated by employing 
NaI(Tl) versus CsI(Tl) scintillators.

%- Apparatus
\subsection{Apparatus}

\begin{figure}[hbt]
 \resizebox{0.45\textwidth}{!}{%
 \includegraphics{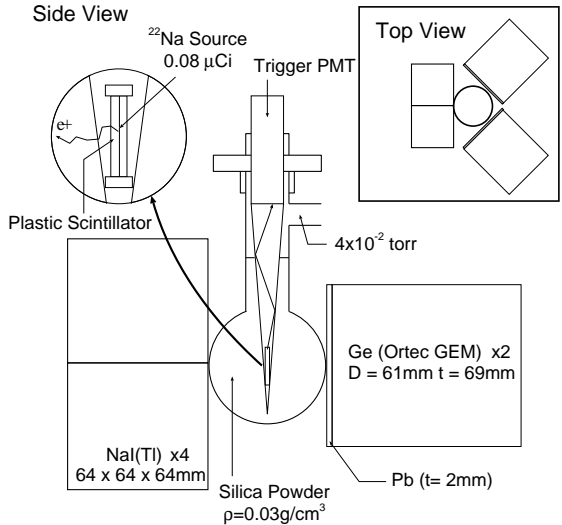}
 }
 \caption{Experimental setup.}
 \label{fig:setup}
\end{figure}

Figure \ref{fig:setup} shows the experimental setup, where a very weak 
${\rm {}^{22}Na}$ $\beta$ positron source ($0.08~\mu Ci$) with 2-$mm$ spot 
diameter is sandwiched between two sheet-type plastic scintillators (NE104) 
having a 12-$mm$ diameter and 100-$\mu m$ thickness. The source, held by a 
cone made of 25-$\mu m$-thick aluminized myler, is situated at the center 
of a 48-$mm$ diameter vacuum chamber made of 500-$\mu m$-thick glass. 
The chamber, evacuated to $4\times10^{-2}$ torr, is filled with 
${\rm SiO_2}$ silica powder, AEROSIL 300CF(obtained from Nippon Aerosil Ltd.),
having a grain size of 7-$nm$ in diameter with density of 0.03 $g/cm^3$.
Since the powder is hydrophile, they were heated for 4 hours at $180^\circ C$ 
to remove the absorbed water molecule.

Most emitted positrons passing through the scintillators transmit a 
light pulse to a trigger photo-multiplier (PMT) (Hamamatsu H-3165-04) 
such that Ps is formed as the positrons are stopped in the silica powder. 
The photons subsequently emitted by the decay of o-Ps are measured using 
two types of detectors. 

Two germanium semiconductor detectors (diameter, 61 $mm$; thickness, 
69 $mm$; Ortec, GEM 38195) precisely determine 
$\lambda_{pick}(t)/\lambda_{3\gamma}$ as a function of time.
 The detectors are arranged perpendicular to each other 
(Fig. \ref{fig:setup}) to reduce the incidence of Compton scattered photons 
from the other detector. A 2.0-$mm$-thick lead sheet located in front of the 
detectors eliminates the contribution of the two simultaneous low-energy 
photons produced by the $3\gamma$-decay of o-Ps. Due to the detector's 
excellent energy resolution, i.e. ,1.78 $keV$ FWHM at 1274.6 $keV$,
the resultant 
energy spectrum allows the monochromatic $2\gamma$ pick-off annihilation to 
be easily distinguished from the continuous $3\gamma$-decay.

The four NaI(Tl) scintillators (crystal size, 64x64x64 $mm$) 
simultaneously obtain time and energy information. 
Because they 
have much higher efficiencies compared to the Ge detectors, while 
also having good time response functions, the obtained time spectrum can 
be fit to Eq. (\ref{eq:population}).

%- Electronics
\subsection{Electronics}
Trigger PMT output is fed to a discriminator which provides the start 
signal for both time-to-digital converters (TDCs) (Hoshin C006 and 
CAEN C414) and the reset signals for other electronics blocked (vetoed) 
during measurements. The TDCs are calibrated by a time-calibrator 
(Ortec 462) having an intrinsic time accuracy of 50 $ppm$.

One output from the Ge detector preamplifier is fed into a 
fast-filter amplifier (FFA)(Ortec 579) whose output is divided into 
three signals by a linear fan-out module (LFF)(Lecroy 428F). 
One signal feeds a constant-fraction discriminator (CFD)(Ortec 473A) 
whose output is used as the stop signal for the TDC, while the other two 
feed a charge-sensitive analogue-to-digital converter (C-ADC) (Lecroy 2249SG).
 This enables measuring energy with a narrow gate width (1.25 $\mu s$) 
and the base-line condition just prior to an event having a 200-$ns$ ADC 
gate width. The differential and integral time of the fast-filter amplifier 
are optimized to obtain good time resolution of 6 $ns$ rms.
To measure a precise energy spectrum, the other output from the 
preamplifier is amplified by a spectroscopy amplifier (Ortec 673) 
such that shaped signals are fed into a peak-holding amplitude-to-digital
converter (PH-ADC) (Hoshin C011) with 40-$\mu s$ gate width.

The output from each of the four NaI(Tl) scintillators is divided 
by the LFF into four signals. One output is fed into a discriminator 
whose output with good time resolution of 1.2 $ns$ rms provides the 
stop signal for the TDCs, while the other three outputs are fed to 
C-ADCs. One C-ADC with 3-$\mu s$ gate width, called the {\it wide ADC} 
(Lecroy 2249W), measures the whole charge for the duration of the 
signal; one with 250-$ns$ gate width, called the {\it narrow ADC} 
(Lecroy 2249SG), measures the charge itself; and one with 180-$ns$ 
gate width, called the {\it base ADC} (Lecroy 2249SG) measures the 
base-line condition of the signal just prior to the event.

Accordingly, each Ge detector and NaI(Tl) scintillator provides 
three types of ADC information. Because pile-up events result in larger 
charge integration on the {\it wide} versus {\it narrow ADC}, their 
values are compared such that these events are effectively rejected. 
The {\it base ADC} is also used to reject pile-up events which can 
occur if the base-line condition is shifted due to any remaining tail 
of the previous signal. All ADCs use entirely independent gate input such 
that they are able to start charge integration using self timing,
which results in the ADC gate width being narrower in accordance with 
signal duration. 

Approximately the o-Ps events were collected for two month,
and during the time a room temperature was maintained at $26.0\pm 0.5^\circ C$ 
to ensure stability of the amplifiers, ADCs, and TDCs.

The absolute strength of magnetic field around the Ps formation assembly
was measured to be $0.5\pm0.1$~Gauss corresponding to a mixing 
between the o-Ps and the p-Ps by the ratio $3\times10^{-11}$. 
Thus the observed o-Ps 
can be considered as a pure o-Ps sample.
\section{Analysis}
Use of the energy spectrum precisely measured by the Ge detectors 
enables determining $\lambda_{pick}(t)/\lambda_{3\gamma}$. 
The energy spectrum of the photon originating from the pure o-Ps 
sample is obtained by subtracting the accidental contribution 
(time range, $2000-3000 ns$) from the measured spectrum.
\begin{figure}[hbt]
 \resizebox{0.45\textwidth}{!}{%
 \includegraphics{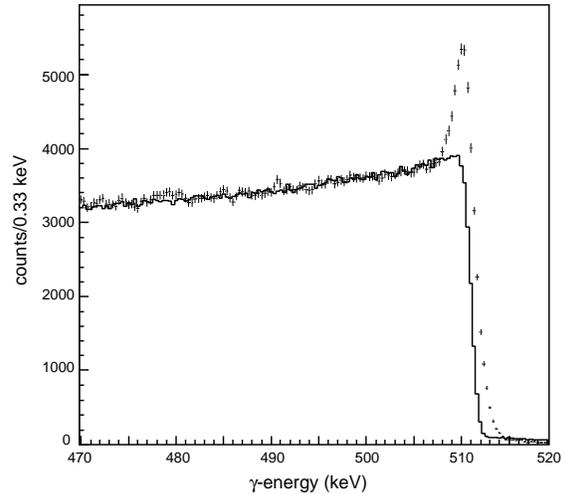}
 }
 \caption{Energy spectrum of o-Ps decay obtained by Ge detectors 
using a time window of $160-710$ ns. The solid line indicates the 
$3\gamma$-decay spectrum calculated by Monte Carlo simulation, 
agreeing well with measured data below 508 $keV$.}
 \label{fig:o-ps}
\end{figure}
\begin{figure}[hbt]
 \resizebox{0.40\textwidth}{!}{%
 \includegraphics{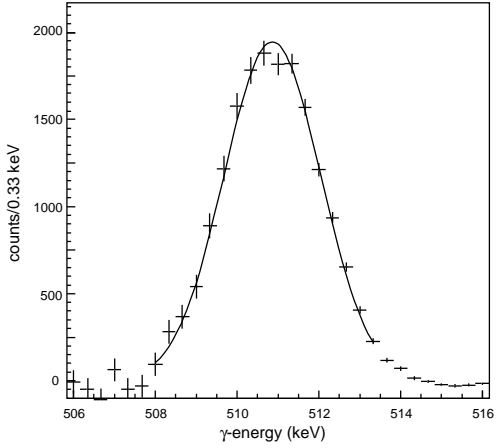}
 }
 \caption{Enlarged view of measured spectrum after subtracting 
the $3\gamma$ contribution from the o-Ps spectrum.}
 \label{fig:pickoff}
\end{figure}
Figure \ref{fig:o-ps} shows the resultant o-Ps spectrum, where the 
pick-off annihilation peak at 511 $keV$ is clearly evident at the edge 
of the $3\gamma$-decay continuous spectrum.

Using Monte Carlo simulation in which the setup geometry and various 
material distributions are reproduced in detail allowed us to 
calculate the expected $3\gamma$-decay energy spectrum. 
That is, for every simulated event, three photons are generated 
according to an $O(\alpha)$ corrected energy spectrum, with 
successive photoelectric, incoherent, or coherent scattering 
interactions of every photon with materials being followed until 
all photon energy is deposited. The response function of the 
detector is determined based on the measured spectrum of 
monochromatic $\gamma$-rays emitted from ${\rm {}^{152}Eu 
(344.3}$ $keV$), ${\rm {}^{85}Sr (514.0}$ $keV$), and ${\rm {}^{137}Cs 
(661.7}$ $keV$), the lines of which are convoluted in the simulation.

As shown in Fig. \ref{fig:o-ps}, the o-Ps and $3\gamma$ spectra 
indicate good agreement below 508 $keV$. Although the simulation 
produced a detailed structure of the actual experimental setup, 
the shape of the $3\gamma$ spectrum is nearly independent of 
simulation details in that the sharp drop-off at 511 $keV$ is solely 
due to the phase-space cutoff, being almost entirely determined 
by the energy response function of the Ge detectors.
Figure \ref{fig:pickoff} shows an enlarged view of the o-Ps 
spectrum after subtracting the $3\gamma$ spectrum, where the 
resultant spectrum peak and measured Ge detector response function 
fit well. Since the center of the fitted peak is located at 
$510.86\pm0.21$ $keV$, this indicates that subtraction of the 
$3\gamma$ contribution is properly performed. Hence, the peak 
is considered to represent a pure sample of pick-off annihilations, 
and the number of pick-off annihilations can be counted.
The number of decays of o-Ps into $3\gamma$'s can accordingly be 
counted using the o-Ps spectrum such that 
$\lambda_{pick}/\lambda_{3\gamma}$ is simply estimated from the 
counts ratio, where $n_{pick}/n_{3\gamma}$ is multiplied by a 
normalization factor considering relative efficiencies of the Ge 
detectors and the $O(\alpha)$ corrected energy spectrum. 
Since $\lambda_{pick}/\lambda_{3\gamma}$ is only dependent on 
the relative efficiencies of the Ge detectors vice absolute efficiencies, 
the resultant ratio is stable against various systematic errors.
%%
% Thermalization %
%%
%
\begin{figure}[hbt]
 \resizebox{0.50\textwidth}{!}{%
 \includegraphics{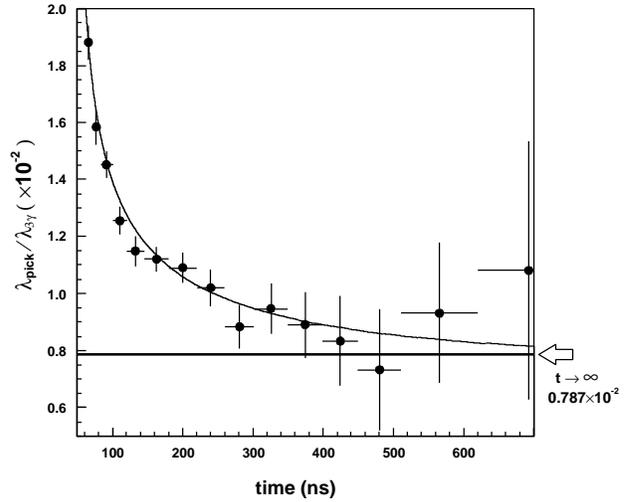}
 }
 \caption{Pick-off annihilation rate and $3\gamma$ decay rate 
plotted as a function of time. Fitting results (solid line) are 
obtained using Eq. (\ref{eq:lambda_ratio}). The horizontal line at 
$\lambda_{pick}/\lambda_{3\gamma}=0.787\times10^{-2}$ 
shows the approximate value at $t\rightarrow\infty$.} 
\label{fig:lambda_ratio}
\end{figure}

$\lambda_{pick}/\lambda_{3\gamma}$ ratios are respectively calculated 
in the same manner using various time windows, with Fig. 
\ref{fig:lambda_ratio} showing time dependencies of these values. 
Because the fractional energy loss of o-Ps per collision is 
dependent on its energy in $\rm{SiO_2}$ powder, the time dependence 
of the average kinetic energy of o-Ps at time $t$, $\avee$, can be 
expressed by the following differential equation \cite{SIO2-1,SIO2-2}:
\begin{eqnarray}
\lefteqn{\frac{d}{dt}\avee=-\sqrt{2m_{Ps}\avee}} \\  
&&\times\left(\avee-\frac{3}{2}k_BT\right)
\left(\frac{2}{\overline{M}~\overline{L}}\right)\sum_{j=0}^{\infty}a_j\left(\frac{\avee}{k_BT}\right)^{j/2}, \nonumber
\end{eqnarray}
where $m_{Ps}, \overline{M}$, and $\overline{L}$ are the mass of o-Ps, 
effective mass at the surface of the $\rm{SiO_2}$ grain, and mean 
distance between the grains.
Since the pick-off annihilation rate is proportional to the average 
velocity of o-Ps, $\lambda_{pick}/\lambda_{3\gamma}(\equiv\theta(t))$ 
can be expressed as
\begin{equation}
\label{eq:lambda_ratio}
\frac{d}{dt}\theta(t)=-C\left(\theta(t)^2-\theta_{\infty}^2\right)\sum_{j=0,1,2,\cdots}
a_j\theta(t)^j,
\end{equation}
where $\theta_{\infty}\equiv\theta(t\rightarrow\infty),\, C$, and 
$a_j(j=1,2,\cdots)$ are constants. With the exception that the 
summation term is replaced with abbreviated form $\theta(t)^\kappa$ 
in which $\kappa$ is an arbitrary real number, Eq. (\ref{eq:lambda_ratio}) 
is used for 
fitting measured $\lambda_{pick}/\lambda_{3\gamma}$ values. 
As indicated by the results in Fig. \ref{fig:lambda_ratio}, o-Ps takes 
about 600 $ns$ 
to become well thermalized in $\rm{SiO_2}$ powder. 
% NaI time spectrum
\begin{figure}[hbt]
 \resizebox{0.45\textwidth}{!}{%
 \includegraphics{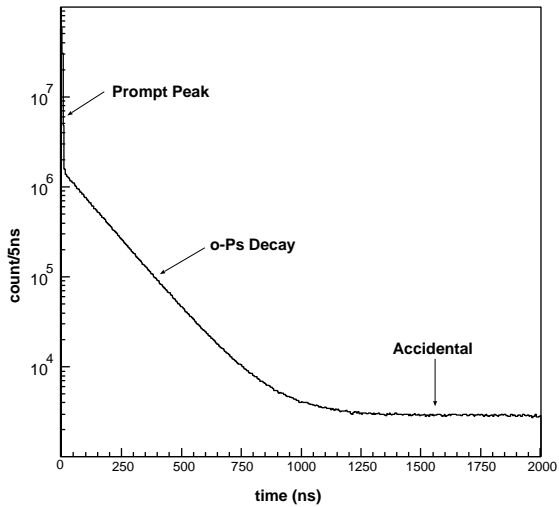}
 }
 \caption{Time spectrum of NaI(Tl) scintillators with a photon energy window of 350-450 $keV$.}
 \label{fig:nai_time}
\end{figure}

Figure \ref{fig:nai_time} shows the time spectrum of NaI(Tl) 
scintillators using a photon energy window of $350-450$ $keV$, where 
the sharp peak from prompt annihilations is followed by exponential 
decay of o-Ps and the flat spectrum of accidental events. 
Due to using a weak positron source ($0.08~\mu Ci$), 
the o-Ps decay curve covers a wide region of about 1.2 $\mu s$ 
, being about eight times as long as the o-Ps lifetime.

Using the least square method the time spectrum is fitted with
\begin{equation}
F(t)=\exp(-R_{stop}t)\left[\left(1+\frac{\epsilon_{pick}}{\epsilon_{3\gamma}}
\frac{\lambda_{pick}}{\lambda_{3\gamma}}\right)N(t)+C\right],
\end{equation}
where $N(t)$ is in Eq. (\ref{eq:population}), and $\epsilon_{pick}$ 
and $\epsilon_{3\gamma}$ are respectively the probabilities that the 
photons emitted from the pick-off annihilation and $3\gamma$-decay 
of o-Ps deposit energy of $350-450$ $keV$ in the NaI(Tl) scintillators. 
Results of Monte Carlo simulation estimated 
$\epsilon_{pick}/\epsilon_{3\gamma}$ to be $0.128\pm0.002$, 
while $R_{stop}$ is the measured stop rate of the NaI(Tl) scintillators.

% decay rate
\begin{figure}[hbt]
 \resizebox{0.45\textwidth}{!}{%
 \includegraphics{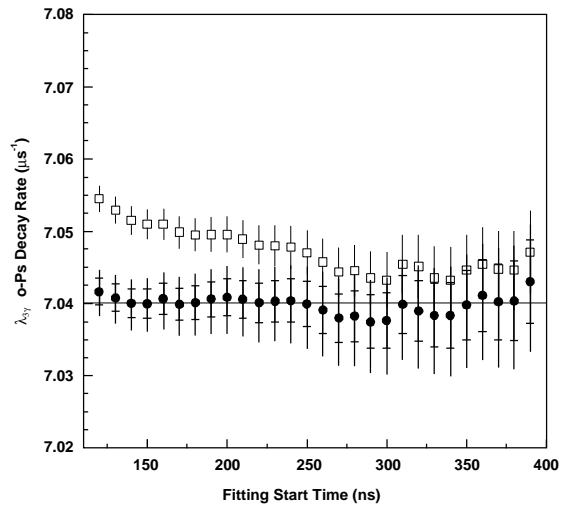}
 }
 \caption{Decay rate of o-Ps plotted as a function of fitting start time. 
Open squares and closed circles show values obtained by fitting 
with Eq. (\ref{eq:simple_expo}) and the proposed method, respectively, 
while the horizontal 
line shows the value obtained at the start of fitting (140 $ns$).}
\label{fig:decay_rate}
\end{figure}
Open squares in Fig. \ref{fig:decay_rate} show fitted values of the decay 
rate of o-Ps, 
$\lambda_{3\gamma}$, plotted as a function of fitting start time 
using a simple exponential function, i.e.,
\begin{equation}
\label{eq:simple_expo}
f(t)=\exp(-R_{stop})\left[~N_0\exp\left(-\lambda_{obs}t\right)+C~\right].
\end{equation}
Since observed $\lambda_{obs}$ contains the pick-off rate fraction, 
the intrinsic decay rate $\lambda_{3\gamma}$ is obtained by 
correcting $\lambda_{obs}$ by a factor of 
$(1+\lambda_{pick}(t\rightarrow\infty)/\lambda_{3\gamma})\approx1.00787$. 
Closed circles in Fig. \ref{fig:decay_rate} show fitting results obtained 
using the 
proposed method, 
where the obtained values are stable against the fitting start time. 
Figure \ref{fig:decay_rate} shows the effect due to thermalization has been 
completely taken into account in our new method.
The horizontal error-bars indicate statistical errors solely due to 
fitting the NaI(Tl) time spectrum, while the vertically extending 
bars represent propagated errors due to the fitting of 
$\lambda_{pick}(t)/\lambda_{3\gamma}$. 
Due to the tail effect of prompt events, fitting $\chi^2$ is rapidly 
increasing before 120 $ns$, instead one can obtain stable $\chi^2$ values
after 140 $ns$. 
Note the stable fitting values after 140 $ns$,  
where the fitting $\chi^2$ is 621.308 for (573-3) degrees of freedom; 
hence this time is used as the fitting start time. 
The resultant value is 
$\lambda_{3\gamma}=7.0401\,^{+0.0033}_{-0.0038}\,(stat.)\,\-\mu s^{-1}$, 
which includes the statistical error in determining 
$\lambda_{pick}(t)/\lambda_{3\gamma}$.

When systematic errors were estimated for various effects, 
the predominate contribution was found to be due the Monte Carlo 
simulation. That is, the $3\gamma$ contribution in the o-Ps spectrum is 
estimated by normalizing the Monte Carlo distribution with measured data 
from $480-505$ $keV$. In this measurement, we used the same Monte Carlo code 
adopted previously \cite{ASAI94}. Equivalent systematic errors originating 
from the simulation are accordingly expected and estimated to be 
$+190$ and $-150$ $ppm$. Other errors arise due to in-uniformity of  
${\rm SiO_2}$ powder ($\pm30~ppm$), fluctuations in relative efficiencies 
of Ge detectors dependent on the position of o-Ps formation ($-40~ppm$), 
and the position of o-Ps decay ($-50~ppm$). The remaining sources of 
error are due to hardware considerations: the absolute accuracy of the 
time calibration ($\pm50~ppm$), uncertainties in the relative 
efficiencies of the Ge detectors ($\pm 50~ppm$), dependence on the 
base-line condition of NaI scintillators ($\pm23~ppm$), and relative 
efficiency of the NaI scintillators ($\epsilon_{pick}/\epsilon_{3\gamma}$) 
($\pm7~ppm$). Since these systematic errors are independent of each other, 
the total systematic error is their quadratic sum.

% decay rate
\begin{figure}[hbt]
 \resizebox{0.45\textwidth}{!}{%
 \includegraphics{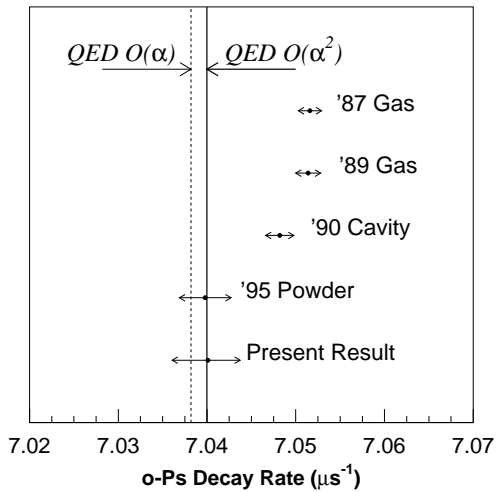}
 }
 \caption{History of the measurement of o-Ps decay rate.
The solid line shows $O(\alpha^2)$ corrected QED prediction, while the dashed line shows $O(\alpha)$ corrected value.}
\label{fig:result}
\end{figure}

The result is $\lambda_{3\gamma}=7.0401\,^{\scriptscriptstyle +0.0033}
_{\scriptscriptstyle -0.0038}\,(stat.)\,^{\scriptscriptstyle +0.0015}
_{\scriptscriptstyle -0.0014}\,\-(sys.)\, \mu s^{-1}$.
Total error is obtained by the quadratic sum of them, i.e.,
$\lambda_{3\gamma}=7.0401\,^{\scriptscriptstyle +0.0036}
_{\scriptscriptstyle -0.0040}$.
Although the obtained precision is relatively lower 
compared to previous experiments, the results are essentially
consistent as shown in Fig. \ref{fig:result} which indicates the history 
of o-Ps decay rate measurements. In addition, substantial improvement 
in the timing system is demonstrated.

\section{Conclusions} 

The o-Ps decay rate was measured in ${\rm SiO_2}$ powder using a 
modified method in which the effect of thermalization of o-Ps is 
accounted for by high energy-resolution Ge detectors. By combining 
obtained and our previous result \cite{ASAI94} that assume 
systematic errors mainly stem 
from the same origins, we obtained 
$\lambda_{3\gamma}=7.0399\,^{\scriptscriptstyle +0.0020}
 _{\scriptscriptstyle -0.0021}\,(stat.) \pm0.0015(sys.)\mu s^{-1}$. 
This is consistent with the $O(\alpha^2)$ corrected QED prediction 
of 7.039934(10)\-$\mu s^{-1}$\cite{ADKINS-4} within an error of 360 ppm, 
while varying $2.7\!\sim\!4.2~\sigma$ from other recent 
measurements \cite{GAS87,GAS89,CAV90}.
Realizing that this experiment represents dynamic development, further 
improvement in Ge timing response is expected soon; an innovation that 
should enable us to investigate the fast component preceeding a 140-$ns$ 
start time such that the $O(\alpha^2)$ contribution in QED can be 
ultimately be confirmed at a level of just 200 $ppm$. 

Sincere gratitude is extended to Professor Toshio Hyodo and Dr. Yasuyuki 
Nagashima, University of Tokyo, for their insightful suggestions and 
valuable discussions concerning o-Ps thermalization.
\bibliographystyle{phreport}

\end{document}